# Preparation of the AlTiNiCuCox system high-entropy alloys and structural analysis

**Junfeng Wang[1], Xinran Wang[1],Zhihao Huang[1], Ting Li[2]，and Yudong Fu[1,*]**

[1]Harbin Engineering University, department of Material Science and Chemical Engineering, Harbin, 150001, China

[2]NorthEast Light Alloy Co., Ltd., Harbin, 150060, China

*fuyudong@hrbeu.edu.cn

## ABSTRACT

This study aimed to explore and develop a new material with high cost-effectiveness, excellent strength, light weight, high hardness, great wear resistance, corrosion resistance, and favorable oxidation resistance. First of all, the $AlTiNiCuCo_X$ system novel high-entropy alloys (HEAs) were prepared in this project according to the vacuum melting method. Subsequently, the metallographical microscopy, XRD diffractometer, scanning electron microscope (SEM), transmission electron microscopy (TEM), and EDS energy spectrum analysis were carried out to observe the microstructures[1, 2] and phase compositions [3-5]of alloys with various Co contents, thus exploring the causes of their numerous favorable performances at structure level.

Structural analysis suggested that, with the change in Co addition amount, the surface morphology and structure of the alloy system changed. XRD analysis indicated that, the alloy system was the FCC+BCC mixed structure. In addition, metallographical demonstrated that, with the increase in Co content, the dendritic crystal transformed from big block to dendritic structure, then to snowflake, gradually to petal-like, and finally to petal shape. SEM-EDS analysis revealed that, Cu element was enriched in interdendritic site, while Ti, Ni, Al and Co elements were enriched in dendrite. Besides, TEM and TEM-EDS analysis indicated that, there was nano-size precipitate of small particles in the Cu-enriched block region, along with dislocation; further, there was twin structure inside the dendrite, as well as the second phase with different morphology, and the second phase showed coherency with the matrix.

The above analysis suggested that, the intercrystalline structure was the Cu-enriched phase of FCC structure; the internal matrix of grain was the NiTi and TiCo phases of BCC structure; and the second phases inside the grain were the $AlCu_2Ti$,$AlNi_2Ti$， $AlCo_2Ti$ and CuNi phases of FCC structure. Taken together, the $AlTiNiCuCo_x$ system novel alloys have changed phase structures and phase types of the alloy system.

## Introduction

The traditional alloy design is associated with limitations and its development tends to be saturated. Scholar Ye Junwei carried out a bold attempt on alloy composition, and proposed a novel alloy in 2004, namely, the multicomponent high-entropy alloy (HEA)[6] . HEA is the alloy that contains at least 5 kinds of elements, and the content of each element is less than 35% (atomic fraction)[7] . Surprisingly, the HEA breaks through the traditional alloy design concept, and substantially broadens the alloy range. Recent research indicates that, numerous HEAs display superb properties, such as high hardness [8], favorable wear resistance [9], temperature resistance [10], strong acidic and alkali corrosion resistance[11, 12], and good oxidative resistance.

Research results reveal that, the multicomponent HEA is the single structure formed by the mutual solid solution of elements at microstructure level[13], most of them form the simple BCC structure and FCC structure, and even the nanocrystal and amorphous phase structure[14, 15], but they can hardly form the complicated while fragile intermetallic compound (IMC) [16]. The feature of high entropy and low enthalpy contributes to forming the solid solution; on the one hand, this is because that the formed degree of order of compound atom arrangement is higher than that of solid solution; on the other hand, the high entropy probably reduces the electronegativity difference, hinders compound formation, contributes to blending of components, and forms the simple face-centered structure and body-centered structure. Besides, according to previous studies, nanophase[17] and even amorphous structure phase precipitation[18-20] is observed in the completely annealed or as cast HEA. The multicomponent HEA displays the nanocrystallization characteristics, which is related to the kinetic theory. To be specific, the atoms are disorderly arranged and at disordered state when the alloy transits from solid state to liquid state. At the time of cooling solidification, the separation of all atoms in multiple elements was blocked, while the atom diffusion of HEA is closely correlated with atom diffusion of all components, the diffusion of component atoms is hindered, and the new phase crystal nucleation and growth are suffocated, thus promoting the formation of small grain nanophase [21]. If the cooling velocity is high enough, the amorphous microstructure will also be generated. The components of HEA are constituted by atoms of various radii, and such atom size differences of the components render lattice distortion around the atoms to various degrees. Actually, the atom radii of HEA components are greatly different from each other, and there is also great lattice distortion effect in the system, which may greatly affect the material properties, such as mechanics, calorifics, and electromagnetism. Each element in HEA is similar to each alcohol in the cocktail, which independently and reciprocally exists in the HEA system. The presence of cocktail effect allows to regulate the HEA structure and performance through controlling the element type and quantity. The above-mentioned effects constitute the four major effects of HEA. At present, the synthetic mechanism of HEA composite material has not been comprehensively understand. Therefore, a large number of novel materials can be synthesized, analyzed and developed, just like the traditional metal-matrix composite materials, so long as the HEA structure is simple and the phase composition can be identified.

To obtain alloy with light weight and superior high temperature oxidation resistance, Ti and Al elements were selected in this study. To improve the strength, Ni, Cu and Co elements were selected, since element Co has high strength under high temperature. HEAs containing Cr or Al can undergo strong acid corrosion, and have superb antioxidation resistance at the temperature of as high as 1100 ℃. In this study, 5 common metal elements, including Ti, Ni, Al, Cu and Co, were selected based on the alloying principle and the electronic structure, and element Co was used as the variable to prepare 5 novel material ingots $AlTiNiCuCo_X$ (x=0.5,1.0,1.5,2.0,2.5) according to vacuum melting method. The new materials obtained through melting were performed microstructure analysis, phase and energy spectrum analyses, so as to screen the novel material with superb properties. This experimental study is of certain reference value in the field of HEA.

## Methods

### Material preparation, alloy structure and phase analysis methods

In this project, 5 common metal elements, including Ti, Ni, Al, Cu and Co, were selected to prepare the HEAs. All the adopted materials were the high-purity alloy elements, with the purity of 99.9%. 5 HEAs were prepared according to the AlTiNiCuCox(x=0.5,1.0,1.5,2.0,2.5) system.

Then, the test specimens were conducted XRD test using the X'Per Pro multifunctional X-diffractometer (PANalytical), at the scanning speed of 4°/min, the scanning angle from 20° to 120°, and the Cu target wavelength λ of 1.54. Then, the corroded test specimens were placed into the OLYMPUS-PM3 optical metallographical microscope, JEOLJSM-6480A scanning electron microscope (SEM) (equipped with energy disperse spectroscope EDS), and JEM-2100 transmission electron microscope (TEM), respectively, to take photos of test specimens at various magnifications. Later, the SEM was used in combination with electron probe for point analysis and regional analysis on the special points and regions in the scanned images, so as to obtain the elemental compositions of corresponding points and regions.

## Results and Discussion

### Morphological analysis

### 3.1 Metallographical structural analysis of the $AlTiNiCuCo_X$ system

Fig.3-1 shows the images of $AlTiNiCuCo_X$ (x=0.5,1.0,1.5,2.0,2.5) system multicomponent HEAs under low-magnification metallographical microscope.

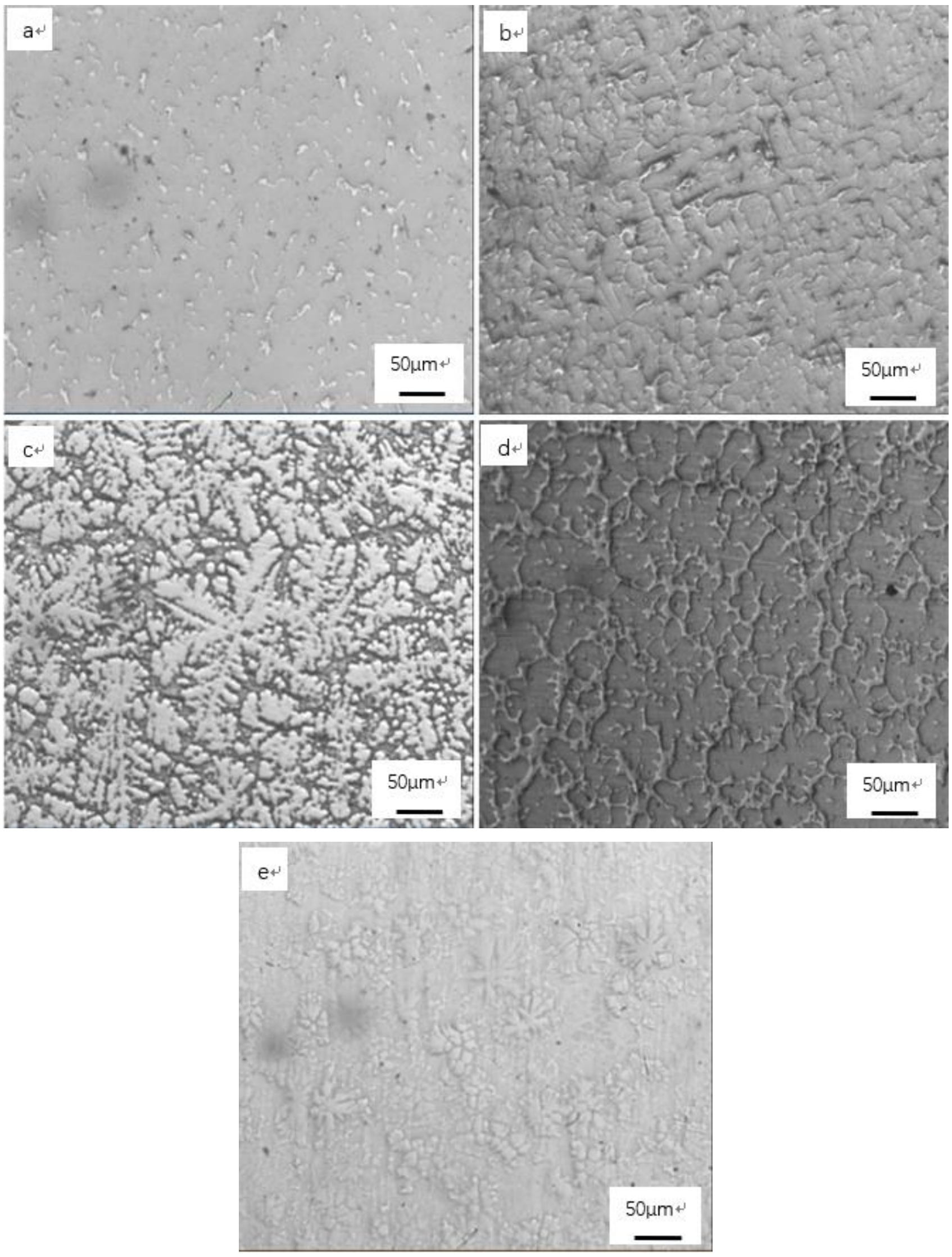

Fig.3-1 Metallographical structures of $AlTiNiCuCo_x$ multicomponent HEAs at low-magnification as cast state

（a）$AlTiNiCuCo_{0.5}$ （b）$AlTiNiCuCo_{1.0}$ （c）$AlTiNiCuCo_{1.5}$

（d）$AlTiNiCuCo_{2.0}$ （e）$AlTiNiCuCo_{2.5}$

Table 3.1 Changes in grain structural morphology, phase and grain boundary at different Co contents

| Element Co content | Grain structural morphology | Phase | Grain boundary |
|---|---|---|---|
| 0.5 | Line and filiform | Light gray phase, gray phase, and little black dot | Relatively clear |
| | | phase | |
| 1.0 | Dendrite-shaped | Dendrite gray-black phase and interdendritic light gray phase | Relatively clear |
| 1.5 | Snowflake-like | Light gray phase and gray black phase | Clear |
| 2.0 | Snowflake-shaped | Gray phase and dark gray black phase | Clear |
| 2.5 | Petal-shaped | Light gray phase and little black point phase | Unclear |

Table 3.1 Changes in grain structural morphology, phase and grain boundary of the $AlTiNiCuCo_x$ ( x =0.5,1.0,1.5,2.0,2.5) system HEAs

The microstructures of the as cast $AlTiNiCuCo_X$ (x=0.5,1.0,1.5,2.0,2.5) system HEAs were analyzed through the metallographical images, the results suggested that, with the increase of Co content, the dendrite shape changed from line and filiform to the regular dendritic shape, then to snowflake-like shape, to petal-like shape, and finally to petal shape gradually. It was observed from the metallographical images that, when the Co content was ≥1.0, the structure showed the obvious 3D perspective, and dendrite projection was observed on the surface. Moreover, with the increase in Co addition amount, the relative area in intergranular region increased, while that in intragranular region decreased, and the grain size gradually became coarse.

**3.2 Phase analysis of the $AlTiNiCuCo_X$ system HEAs**

Fig.3-2 shows the XRD diagram of the $AlTiNiCuCo_X$ system HEAs

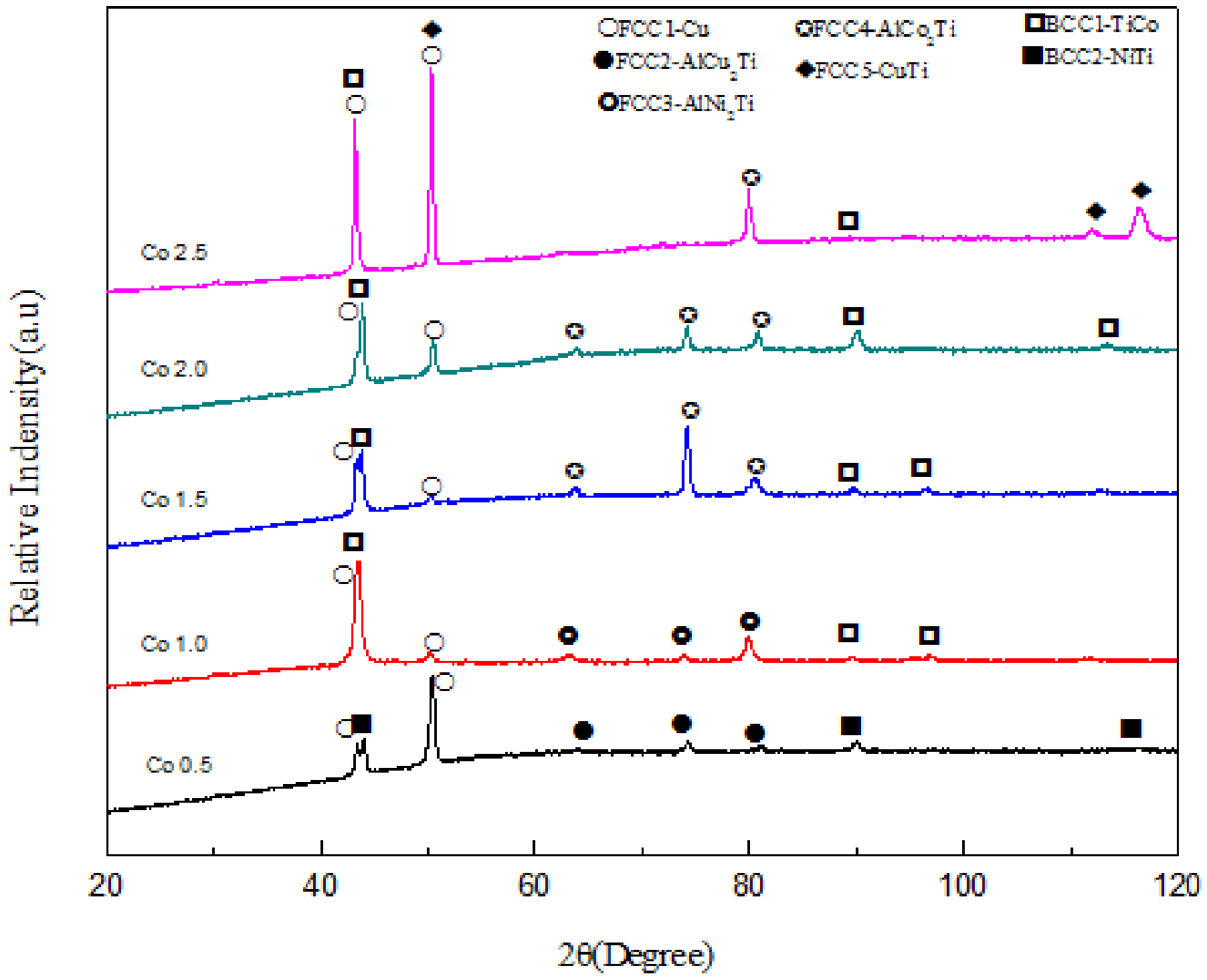

(a) x=0.5,(b) x=1.0,(c) x=1.5,(d) x=2.0,(e) x=2.5

Fig.3-2 The XRD diagram of the AlTiNiCuCoX system HEAs

Table 3.2 Phase compositions at different Co contents

| Co concentration | FCC | BCC |
|---|---|---|
| 0.5 | FCC1: Cu-rich phase<br>FCC2: $AlCu_2Ti$ phase | BCC2: NiTi phase |
| 1.0 | FCC1: Cu-rich phase<br>FCC3:$AlNi_2Ti$ phase | BCC1: TiCo phase |
| 1.5 | FCC1: Cu-rich phase,<br>FCC4: $AlCo_2Ti$ phase | BCC1: TiCo phase |
| 2.0 | FCC1: Cu-rich phase,<br>FCC4: $AlCo_2Ti$ phase | BCC1: TiCo phase |
| 2.5 | FCC1: Cu-rich phase, FCC5: CuTi phase | BCC1: TiCo phase |

Table 3.2 Different phase compositions of the $AlTiNiCuCo_x$（x ＝0.5,1.0,1.5,2.0,2.5）system HEAs

With the increase in Co content, the phase types in FCC and BCC structures of the alloy system changed. For FCC structure, the Cu phase existed all the time, while the $AlCu_2Ti$ phase gradually disappeared, $AlNi_2Ti$ and $AlCo_2Ti$ phases appeared, and the CuTi phase also appeared in the $AlTiNiCuCo_{2.5}$ alloy. For the BCC structure, the NiTi phase disappeared, and the new TiCo phase appeared.

### 3.3 Scanned images analysis of $AlTiNiCuCo_x$ system HEAs

Fig.3-3 displays the SEM images on the surface structure of the as cast $AlTiNiCuCo_x$ (x=0.5,1.0,1.5,2.0,2.5) system HEAs.

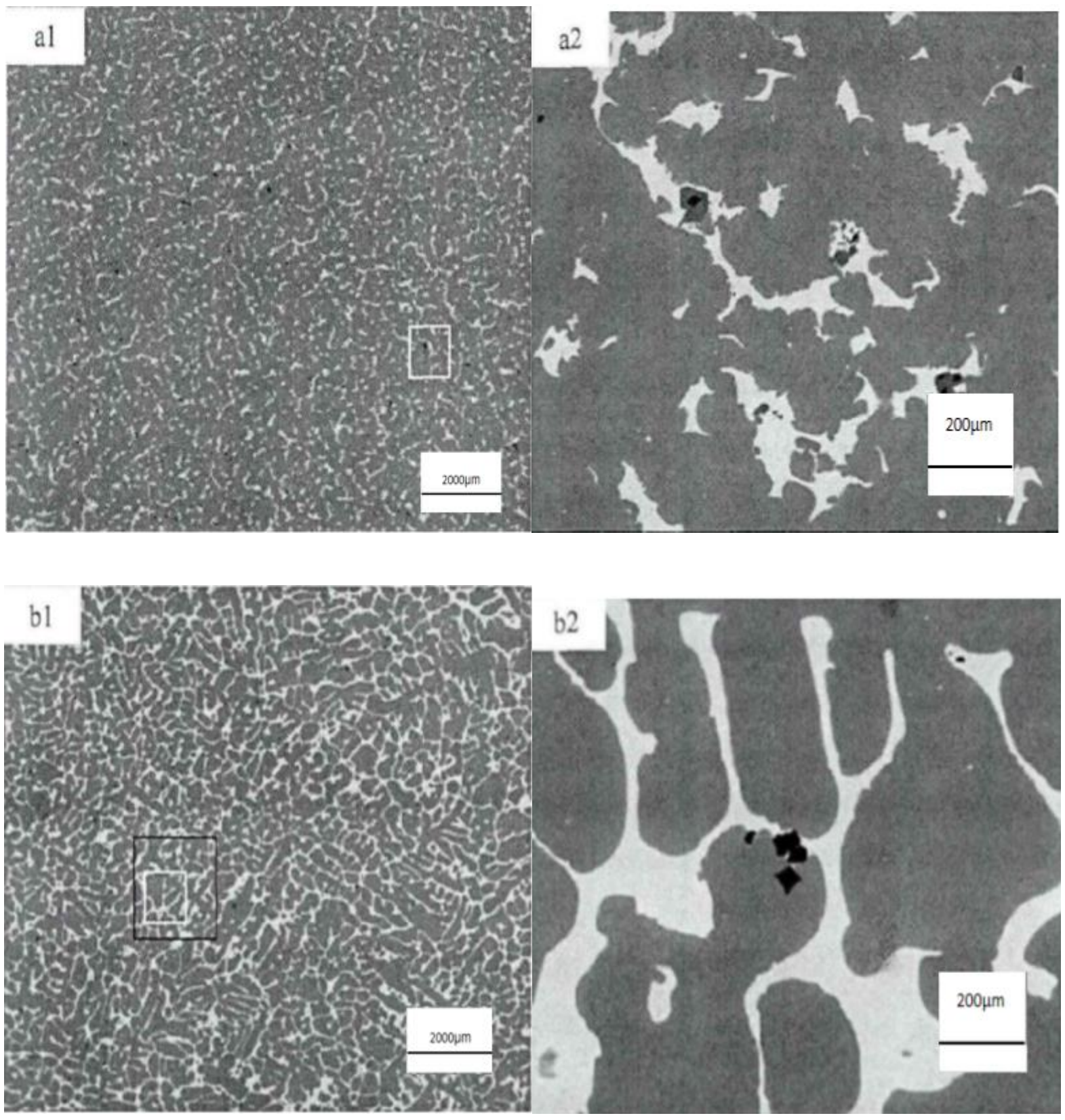

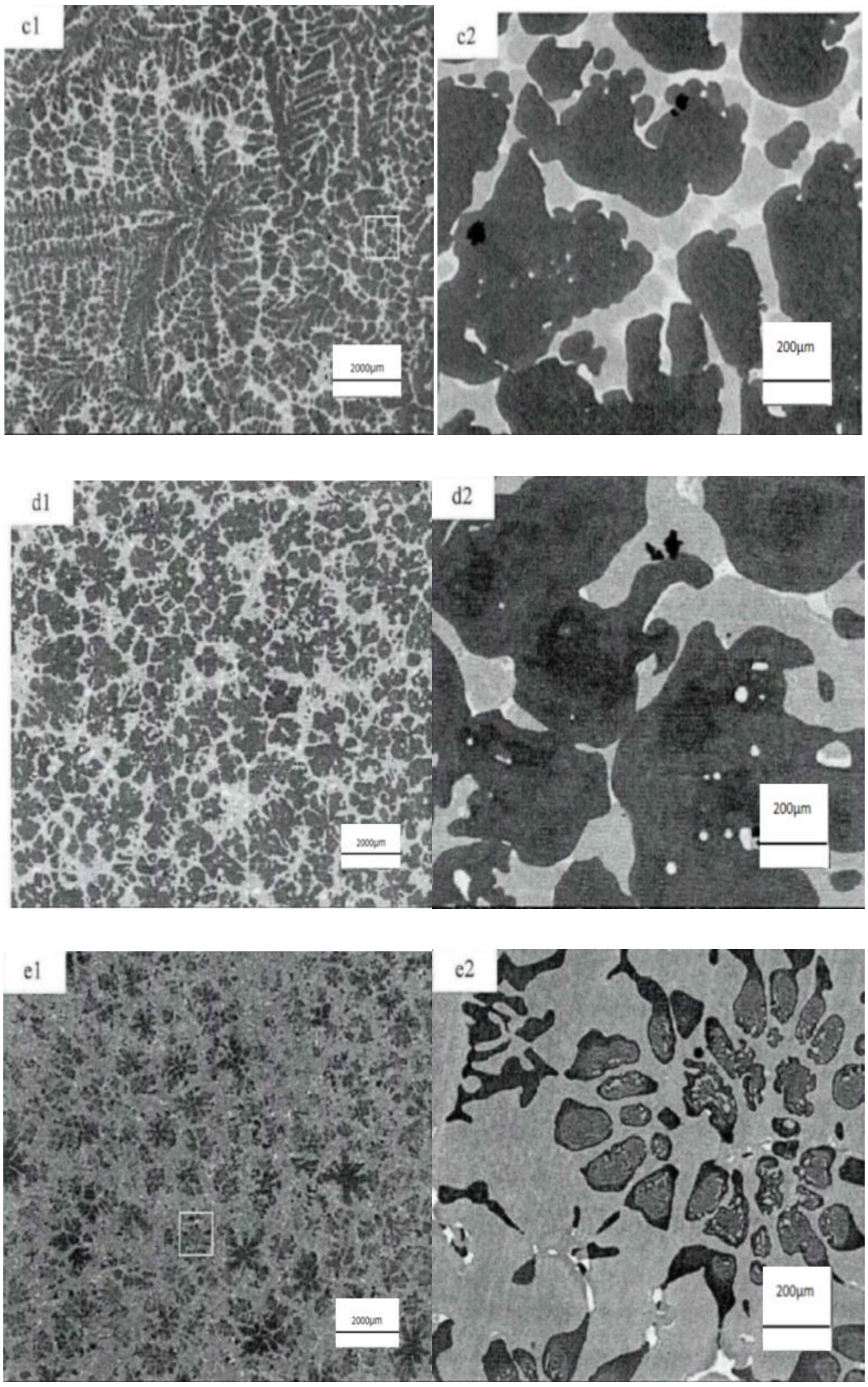

Fig.3-3 SEM images of the as cast $AlTiNiCuCo_x$ system HEAs at 100 and 1000 magnifications □

(a) x=0.5，(b)x=1.0，(c)x=1.5，(d)x=2.0,(e)x=2.5□□

Table 3.3 Grain boundary, phase and phase distribution at different Co contents

| Co content | Grain boundary | Phase | Phase distribution |
|---|---|---|---|
| 0.5 | Relatively | Light gray | The light gray |
| | clear grain boundary | phase, gray phase, little black dot phase | phase displayed the line and silk shape, the little black dot phase showed diffuse distribution, the gray phase exhibited stripe shape (the stripes were interleaved horizontally and vertically) |
| 1.0 | More obvious grain boundary | little black dot phase, gray phase, light gray phase | The little black dot phase showed diffuse distribution, the gray phase began to be regular and showed independent grain shape, the light gray phase no longer |
| | | | displayed the line and silk shape, but the intergranular shape |
| 1.5 | Blurred grain boundary | Little black dot phase, gray phase, light gray phase, white phase | The black little dot phase remained diffuse distribution |
| 2.0 | More blurred grain boundary | Little black dot phase, gray phase, light gray phase, white phase | The diffusedly distributed blank little dot phase, gray phase, light gray phase and white phase |
| 2.5 | Blurred grain boundary | Little black dot phase, gray phase, light gray phase, white phase | The black phase increased inside the structure |

Table 3.3 Grain boundary, phase and phase distribution of the AlTiNiCuCo$_x$ （x ＝0.5,1.0,1.5,2.0,2.5）system HEAs

The scanned images of the as cast AlTiNiCuCo$_x$ (x=0.5,1.0,1.5,2.0,2.5) system HEAs at low magnification were observed on the whole, so as to observe the microstructure and morphology variation rules on alloy surface. To be specific, with the increase in Co content, the variation rules of the alloy surface microstructure and morphology were that, the grains were first refined, then the dendrite arm length became shorter, the width became wider, and the filiform shape changed into the regular dendrite crystal structure. With the further increase in Co, the dendrite crystal structure changed into the coconut branch shape, and then gradually into the petal shape dendrite crystal structure. Besides, it was seen from the locally magnified scanned images of the system $AlTiNiCuCo_x$ HEAs that, with the increase in Co addition amount, the black little dot phase gradually decreased; the contrast began to occur in the microstructure when the Co addition amount (molar ratio) was >1.5, and the contrast was not obvious at the Co addition amount of 2.5, with the production of the block black phase.

### 3.4 EDS energy spectrum analysis for the $AlTiNiCuCo_x$ system HEAs

Fig.3-4 and Table 3.4 display the SEM-EDS diagram of $AlTiNiCuCo_{0.5}$ alloy surface microstructure under 1000 magnification and the table of results of EDS analysis of surface area of abrasion marks.

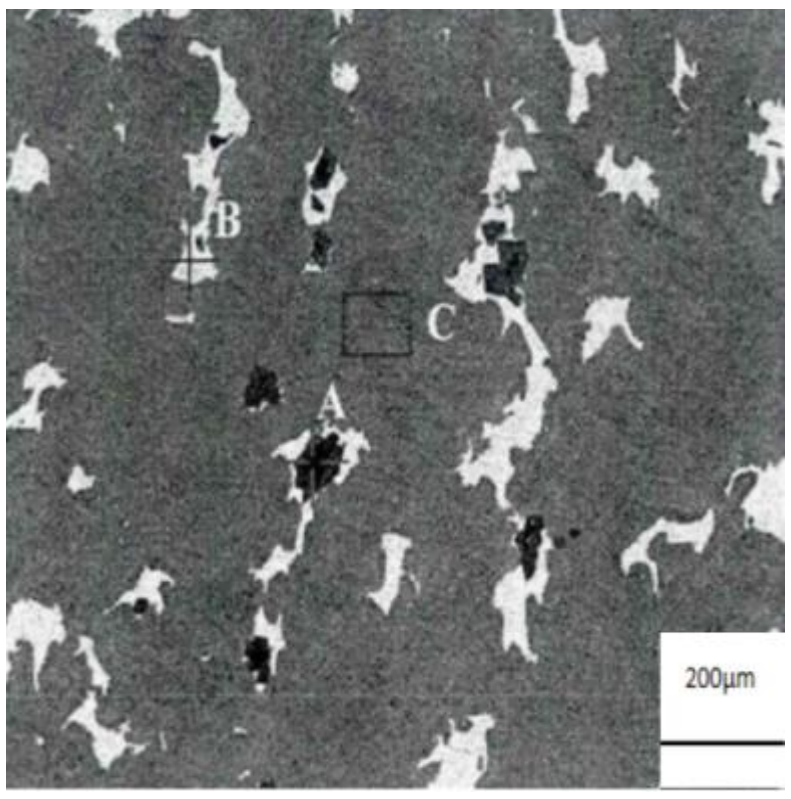

Fig.3-4 Scanned ESD diagram of $AlTiNiCuCo_{0.5}$ as cast alloy surface microstructure

A, black dot part, B, light gray part, C gray part

| Position | A | B | C |
|---|---|---|---|
| Ti | 20.58% | 1.54% | 22.83% |
| Ni | 2.11% | 2.44% | 27.35% |
| Al | 72.65% | 11.02% | 23.34% |
| Cu | 2.71% | 83.52% | 11.77% |
| Co | 1.95% | 1.47% | 14.71% |

Table 3.4 EDS analysis results of surface area of AlTiNiCuCo $_{0.5}$ abrasion mark

Fig.3-5 and Table 3.5 present the SEM-EDS diagram of AlTiNiCuCo$_{1.0}$ alloy surface microstructure under 1000 magnification and the table of results of EDS analysis of surface area of abrasion marks.

□□

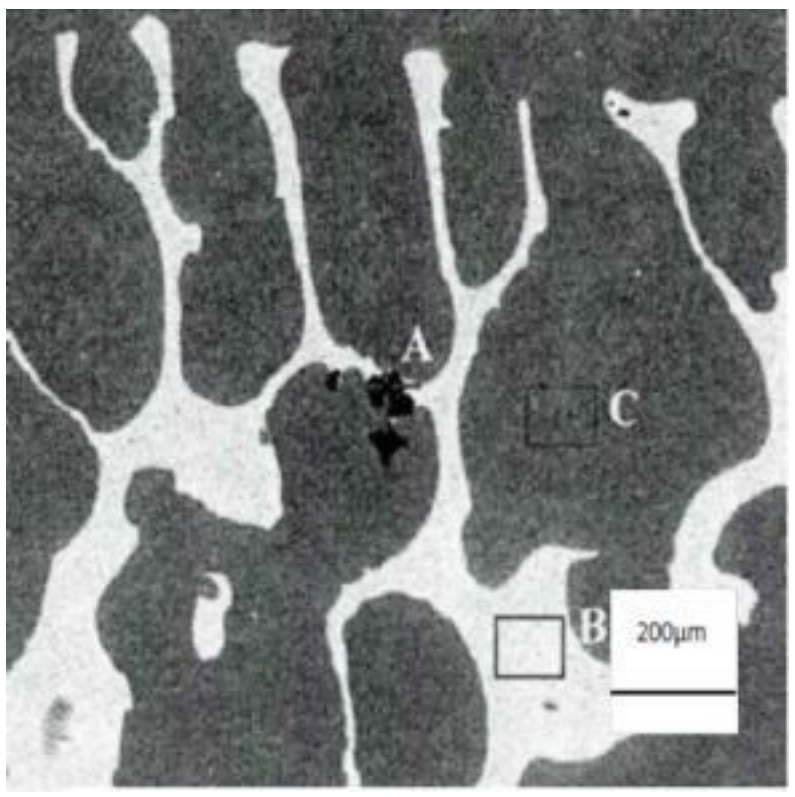

Fig.3-5 Scanned EDS diagram of AlTiNiCuCo$_{1.0}$ as cast alloy surface microstructure

A, black dot part, B, light gray part, C gray part

| Position | A | B | C |
|---|---|---|---|
| Ti | 54.97% | 2.65% | 22.91% |
| Ni | 1.43% | 12.40% | 21.51% |
| Al | 39.98% | 7.65% | 24.40% |
| Cu | 2.06% | 72.96% | 6.36% |
| Co | 1.55% | 4.35% | 24.83% |

Table 3.5 EDS analysis results of surface area of AlTiNiCuCo $_{1.0}$ abrasion mark

Fig.3-6 and Table 3.6 exhibit the SEM-EDS diagram of AlTiNiCuCo$_{1.5}$ alloy surface microstructure under 1000 magnification and the table of results of EDS analysis of surface area of abrasion marks.

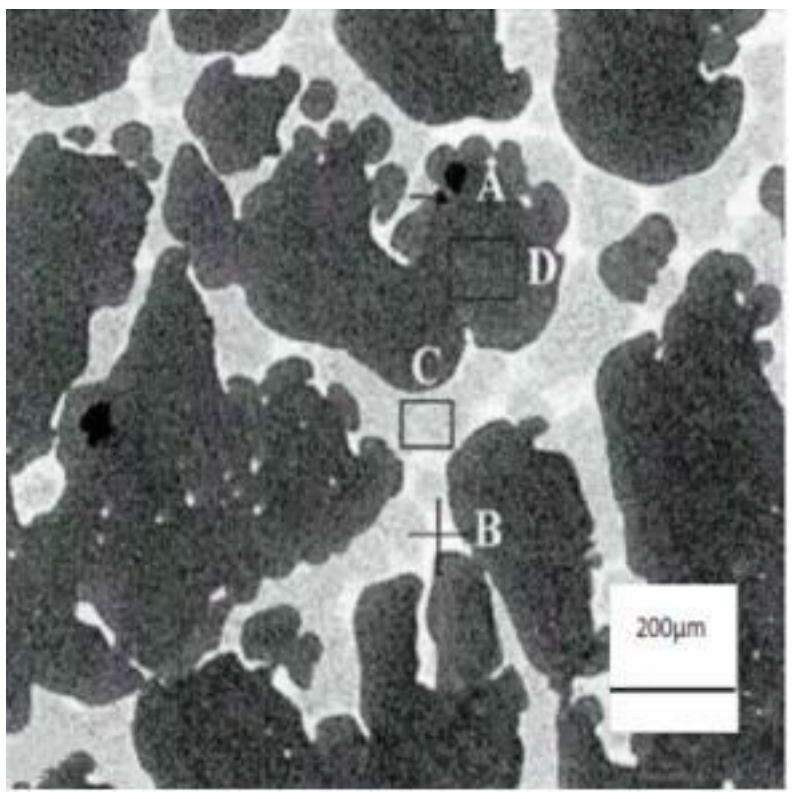

Fig.3-6 Scanned EDS diagram of $AlTiNiCuCo_{1.5}$ as cast alloy surface microstructure

A, black dot part, B, white part, C, light gray part, C gray part

| Position | A | B | C | D |
|---|---|---|---|---|
| Ti | 29.22% | 2.52% | 10.71% | 21.52% |
| Ni | 14.02% | 11.77% | 22.39% | 18.13% |
| Al | 31.27% | 6.76% | 8.49% | 23.26% |
| Cu | 5.04% | 71.38% | 35.53% | 4.55% |
| Co | 20.45% | 7.57% | 22.88% | 32.55% |

Table 3.6 EDS analysis results of surface area of $AlTiNiCuCo_{1.5}$ abrasion mark

Fig.3-7 and Table 3.7 exhibit the SEM-EDS diagram of $AlTiNiCuCo_{2.0}$ alloy surface microstructure under 1000 magnification and the table of results of EDS analysis of surface area of abrasion marks.

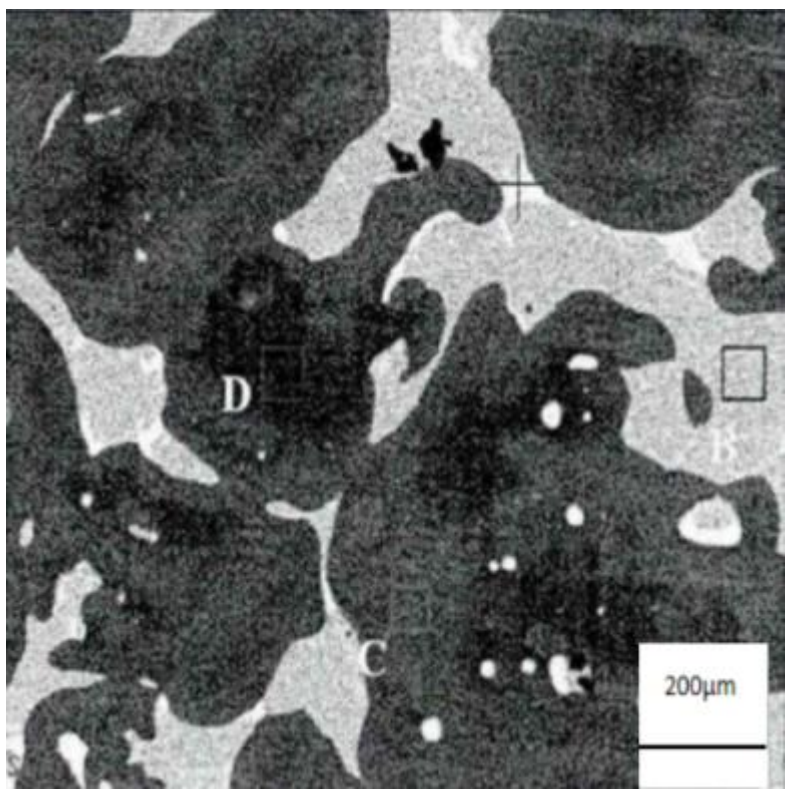

Fig.3-7 Scanned EDS diagram of $AlTiNiCuCo_{2.0}$ as cast alloy surface microstructure

A, white part, B, light gray part, C, dark gray part, D black-wrapped part

| Position | A | B | C | D |
|---|---|---|---|---|
| Ti | 1.67% | 9.13% | 18.32% | 18.75% |
| Ni | 8.01% | 19.65% | 17.43% | 11.68% |
| Al | 7.78% | 9.57% | 20.67% | 32.59% |
| Cu | 75.87% | 29.26% | 5.61% | 4.59% |
| Co | 6.66% | 32.39% | 37.96% | 32.40% |

Table 3.7 EDS analysis results of surface area of $AlTiNiCuCo_{2.0}$ abrasion mark

Fig.3-8 and Table 3.8 exhibit the SEM-EDS diagram of $AlTiNiCuCo_{2.5}$ alloy surface microstructure under 1000 magnification and the table of results of EDS analysis of surface area of abrasion marks.

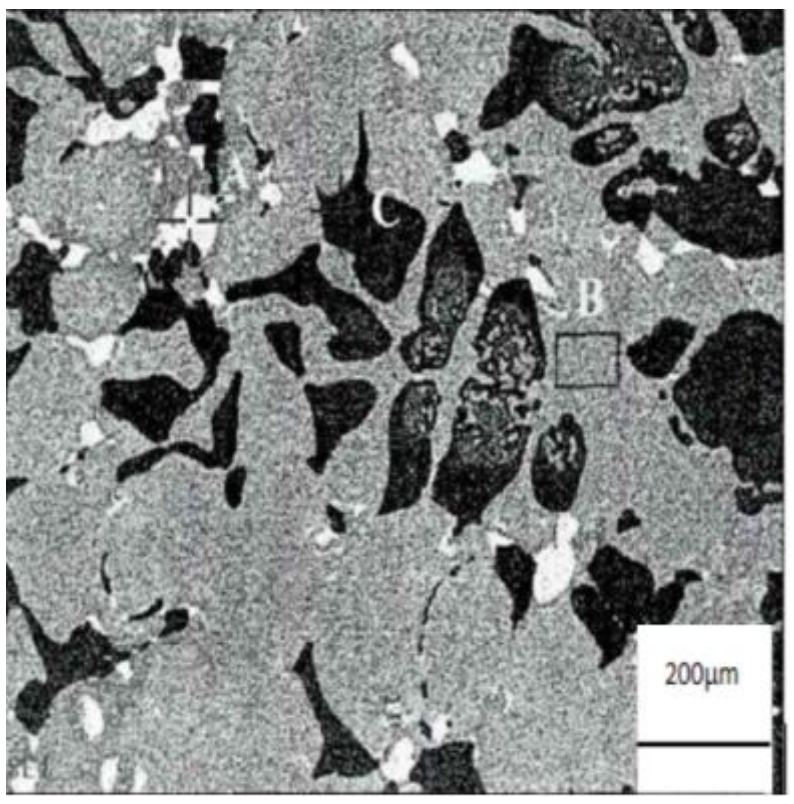

Fig.3-8 Scanned EDS diagram of $AlTiNiCuCo_{2.5}$ as cast alloy surface microstructure

A, white part, B, gray part, C, black part

| Position | A | B | C |
|---|---|---|---|
| Ti | 1.83% | 9.35% | 19.59% |
| Ni | 6.70% | 19.15% | 15.74% |
| Al | 5.22% | 9.87% | 19.63% |
| Cu | 77.90% | 26.34% | 4.05% |
| Co | 8.35% | 35.28% | 41.00% |

Table 3.8 EDS analysis results of surface area of $AlTiNiCuCo_{2.5}$abrasion mark

It was observed from the above tables and diagrams that, the microstructural composition distribution of the as cast multicomponent $AlTiNiCuCo_x$ (x=0.5,1.0,1.5,2.0,2.5) HEAs exhibited the following rules: (1) with the increase in Co content, the intergranular Cu enrichment amount gradually decreased and then increased, the lowest Cu enrichment amount was achieved at the Co molar ratio of 1.5, and the intergranular site was preliminarily speculated to be the Cu-rich phase of FCC structure. (2) With the increase in Co content, the diffusely distributed black dot phase gradually decreased, and structural phase areas of three major regions occurred in the microstructure at the Co contents of 1.5 and 2.0, in addition to the diffusely distributed black dot phase. (3) At the Co content of 2.5, there was a black phase wrapped within the gray phase region in the structure.

Table 3.9 shows the areas where each phase locates at different Co contents

| Co content | FCC: Cu-rich phase | FCC2: $AlCu_2Ti$ phase | FCC: $AlNi_2Ti$ phase | FCC3:$AlCo_2Ti$ phase | FCC: CuTi phase | BCC2: NiTi phase | BCC: TiCo phase |
|---|---|---|---|---|---|---|---|
| 0.5 | B | A | | | | C | |
| 1.0 | B | | A | | | | C |
| 1.5 | B | | | A | | | C,D |
| 2.0 | A | | | D | | | B,C |
| 2.5 | A | | | | B | | C |

Table 3.9 Area where each phase of the $AlTiNiCuCo_x$（ x ＝0.5,1.0,1.5,2.0,2.5）system HEAs locates

## 3.5 TEM image analysis of the AlTiNiCuCoX system HEAs

Fig.3-9 shows the TEM light field images of the as cast $AlTiNiCuCo_{0.5}$ alloy at various regions and the diffraction spots in corresponding areas

Fig.3-9(b) shows the local magnified image of Cu-rich area A. Fig.3-9(c) displays the local magnified image of grain area B

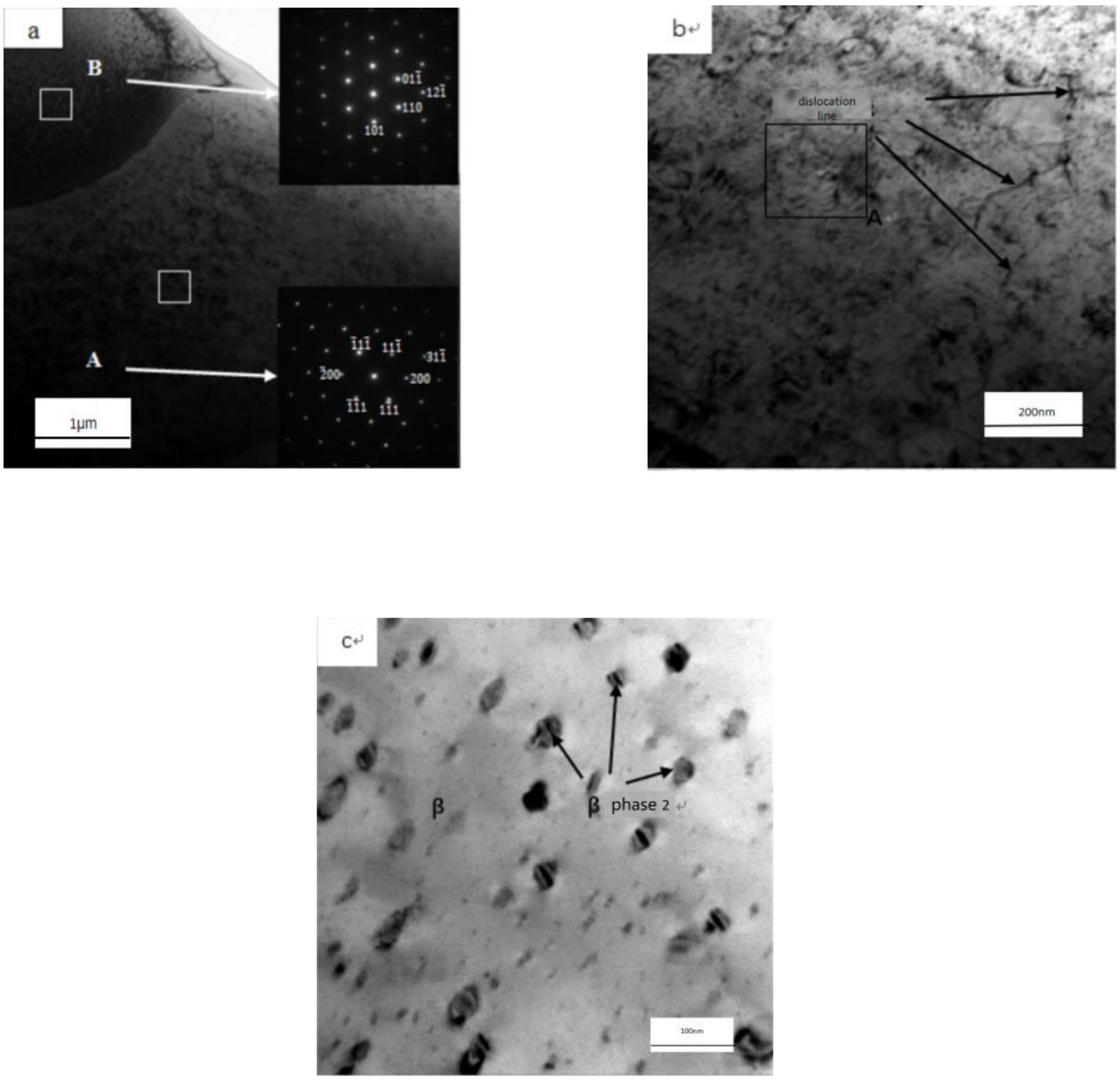

Fig.3-9 TEM analysis results of the as cast $AlTiNiCuCo_{0.5}$ alloy at various regions. (a) TEM morphology at different regions and the diffraction spots in corresponding areas (A-intergranular, B-grain area), (b) magnified TEM morphology of area A in Fig.(a); (c) magnified TEM morphology of area B in Fig.(a).

Fig.3-10 shows the TEM light field image of the as cast $AlTiNiCuCo_{1.0}$ alloy at various regions and the diffraction spots in corresponding areas.

Fig.3-10(b) shows the magnified image of Cu-rich area A. Fig.3-10(c) displays the magnified image of grain area B.

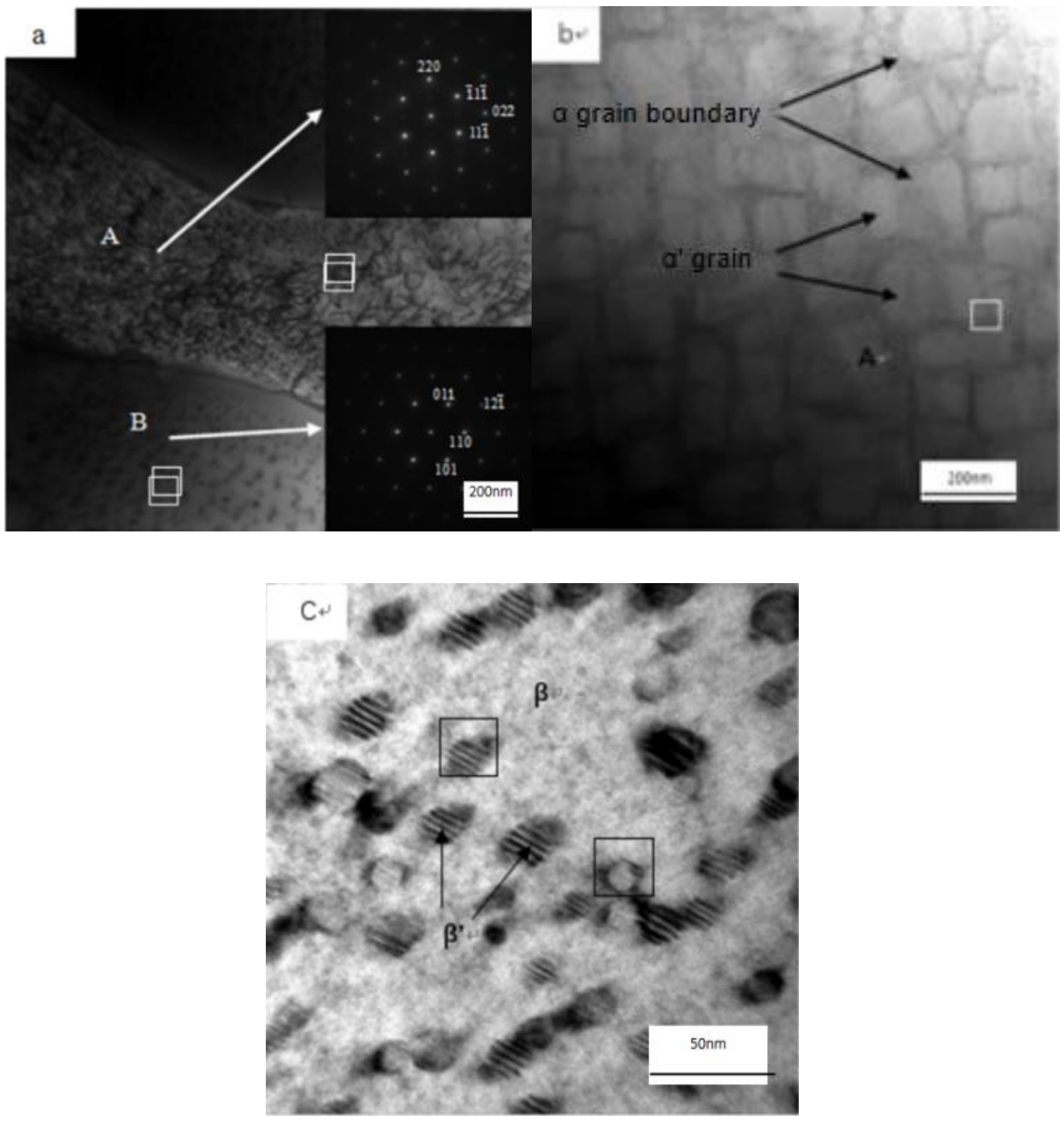

Fig.3-10 TEM analysis results of the as cast $AlTiNiCuCo_{1.0}$ alloy at various regions. (a) TEM morphology at different regions and the diffraction spots in corresponding areas (A-intergranular, B-grain area), (b) magnified TEM morphology of area A in Fig.(a); (c) magnified TEM morphology of area B in Fig.(a).

Fig.3-11 shows the TEM light field image of the as cast $AlTiNiCuCo_{15}$ alloy at various regions and the diffraction spots in corresponding areas. Fig.3-11(b) shows the electron diffraction spots of area A along the zone axis [0001]. Fig.3-11(c) displays the high-resolution TEM image of area A.

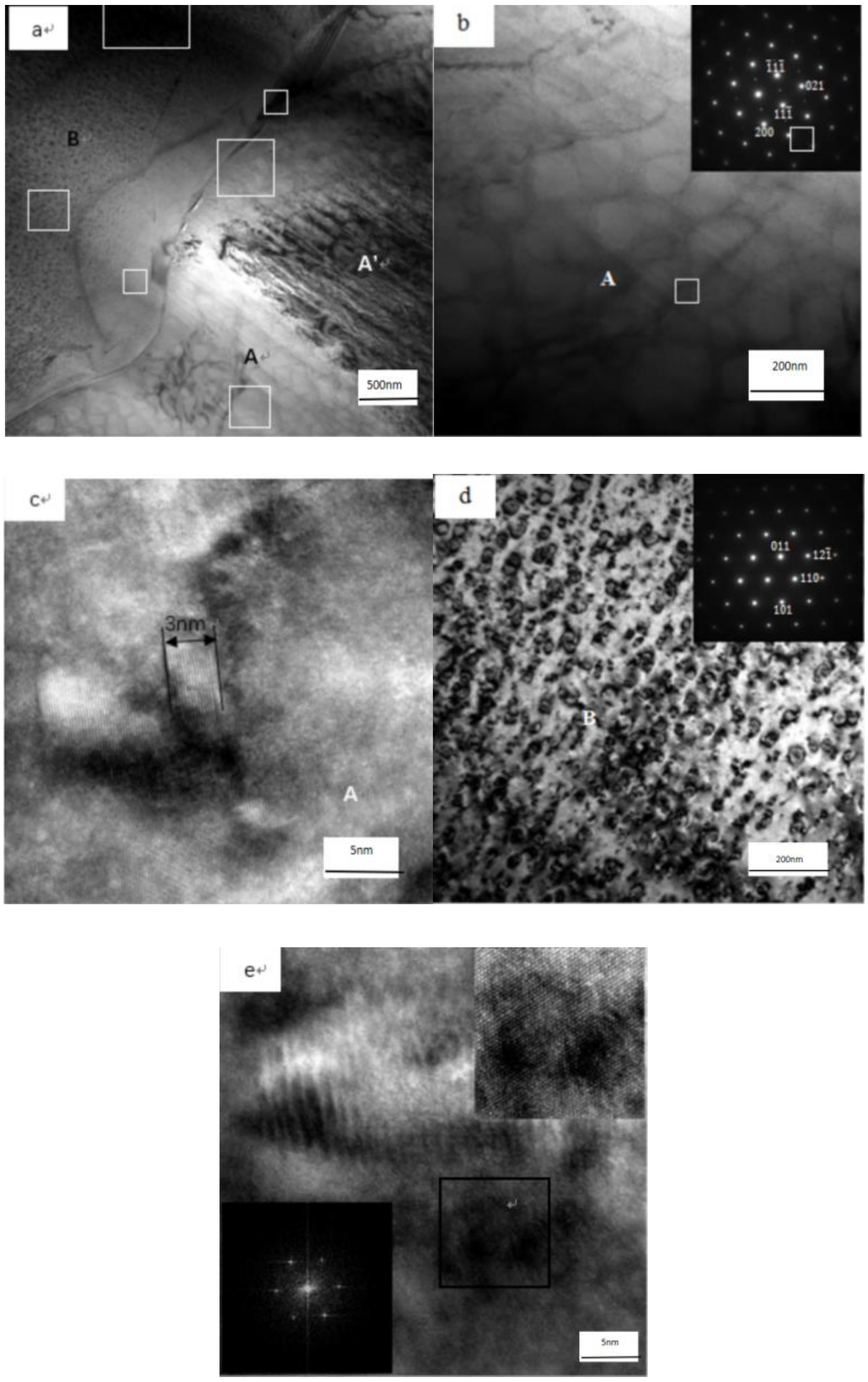

Fig.3-11 TEM analysis results of the as cast $AlTiNiCuCo_{1.5}$ alloy at various regions. (a) TEM morphology at different regions and the diffraction spots in corresponding areas (A-intergranular, B-grain area), (b) magnified TEM morphology of area A; (b) magnified TEM morphology in area A; (c) high resolution TEM image of area A, (d) magnified TEM morphology of area B; (e) high resolution TEM image of area B

Fig.3-12 displays the TEM light field images of the as cast $AlTiNiCuCo_{2.0}$ alloy at various regions and the diffraction spots in corresponding areas. Fig.3-12(b) shows the electron diffraction spots of area B along the zone axis [-111]. Fig.3-12(c) displays the local high-resolution TEM image of grain B.

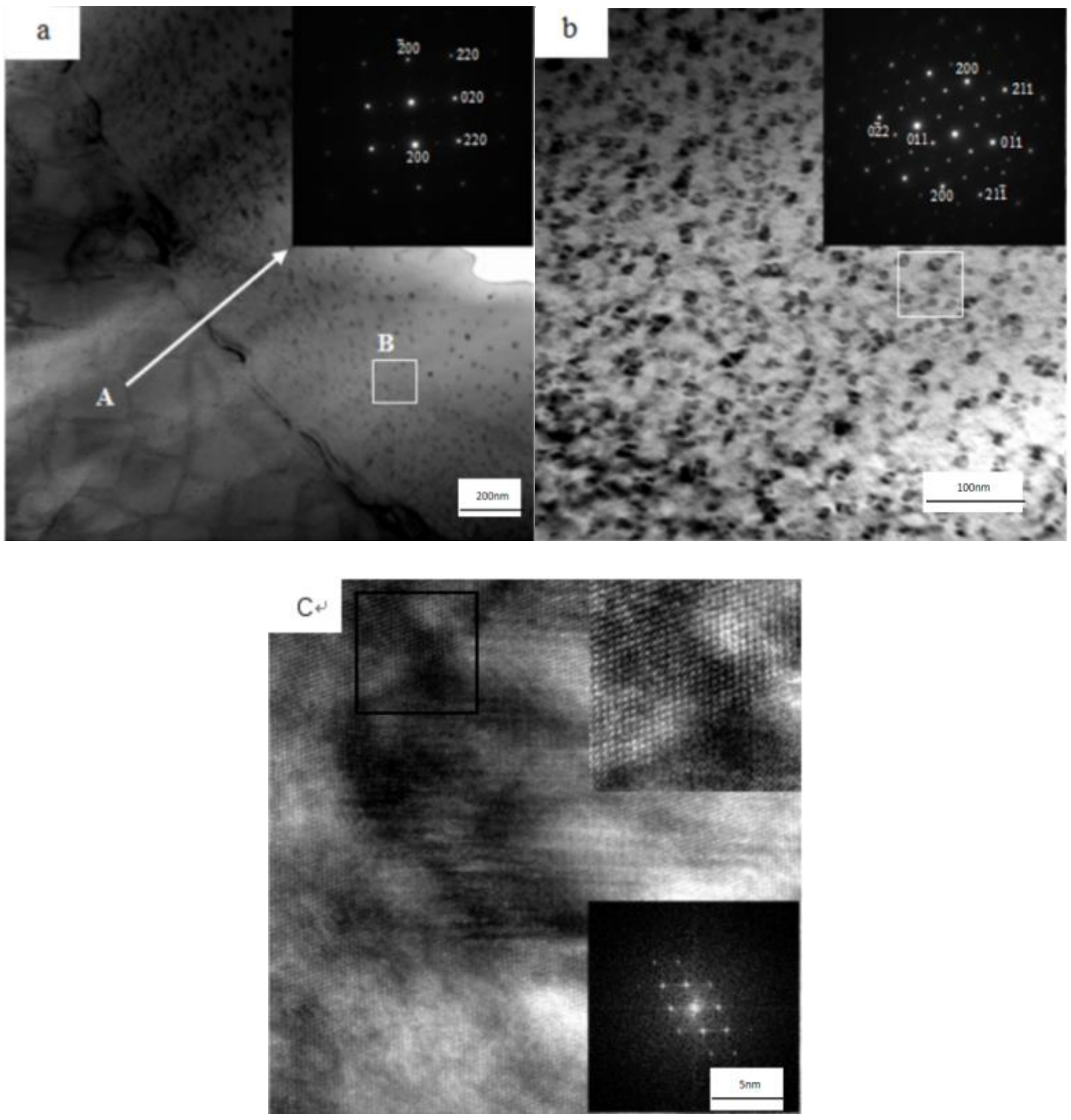

Fig.3-12 TEM analysis results of the as cast AlTiNiCu$Co_{2.0}$ alloy at various regions. (a) TEM morphology at different regions and the diffraction spots in corresponding areas (A-intergranular, B-grain area), (b) magnified TEM morphology of area A; (b) high resolution TEM image of area B

Fig.3-13 displays the TEM light field image of the as cast AlTiNiCu$Co_{2.5}$ alloy at various regions and the diffraction spots in corresponding areas. Fig.3-13(a) shows the electron diffraction pattern of area A along the zone axis [001]. Fig.3-13(b) displays the TEM image of area B and diffraction pattern in selected area.

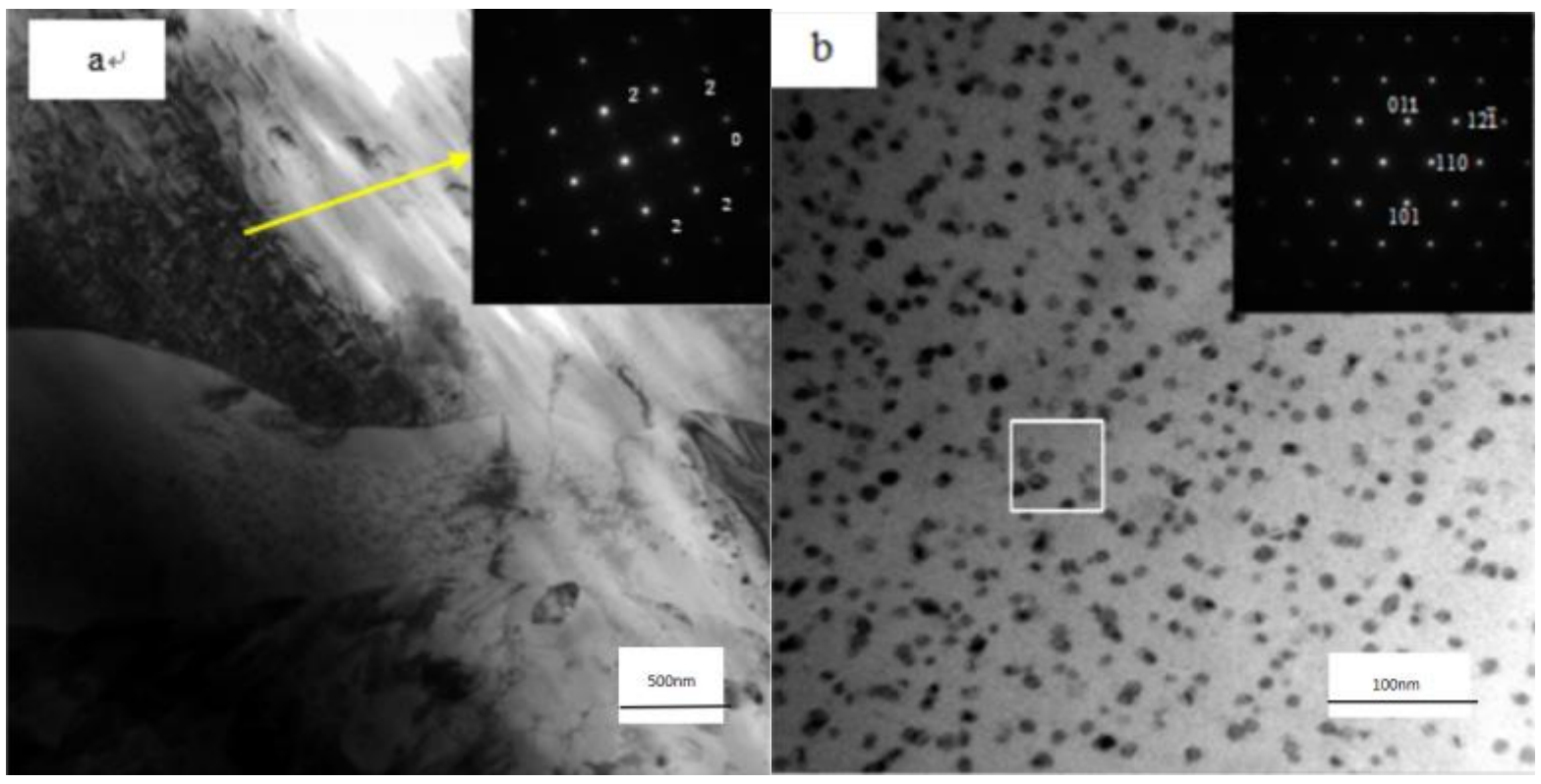

Fig.3-13 TEM analysis results of the as cast AlTiNiCu$Co_{2.5}$ alloy at various regions. (a) TEM morphology at different regions and the diffraction spots in corresponding areas (A-intergranular, B-grain area), (b) magnified TEM morphology of area

A

Table 3.5 shows different Phase distributions, second phase sizes and combining modes at different Co contents

| Co content | Area A | Matrix in area B | Second phase in area B | Second phase size | Combining mode |
|---|---|---|---|---|---|
| 0.5 | Cu-rich phase | NiTi phase | $AlCu_2Ti$ phase | 33nm | Coherency |
| 1.0 | Cu-rich phase | TiCo phase | $AlNi_2Ti$ phase | 17nm | Coherency |
| 1.5 | Cu-rich phase | TiCo phase | $AlCo_2Ti$ phase | 20nm | Coherency |
| 2.0 | Cu-rich phase | TiCo phase | Cu-rich phase | 15nm | Coherency |
| 2.5 | Cu-rich phase | TiCo phase | CuTi phase | 7nm | Coherency |

Table 3.5 Phase distribution, second phase size and combining mode of the $AlTiNiCuCo_x$（ x ＝0.5,1.0,1.5,2.0,2.5）system HEAs

There was second phase precipitation inside the grains, which showed coherency with the matrix. With the increase in Co content, the second phase shape in dendrite regions rich in Ti, Ni, Al and Co elements changed from the multilayer black alternating with white elliptical shape to the multilayer black alternating with white round shape, then to the black granular shape, the number of light alternating with dark layer at second phase first increased and then decreased, and finally turned into the black phase. The second phase size first decreased, then increased, and then decreased; the intragranular second phase of $Co_{2.5}$ was the CuNi phase, while those in remaining alloys were the close $AlNi_2Ti$, $AlCu_2Ti$ and $AlCo_2Ti$ phases. With the increase in Co content, the dislocation lines gradually increased and formed the dislocation network, while the dislocation was generated due to the stress during the alloy cooling process.

## Conclusions

The effect of Co element on the microstructure and phase composition of the alloy was investigated by metallographic microscope, scanning microscope, transmission electron microscope, EDS and XRD analysis in a multicomponent high-entropy alloy $AlTiNiCuCo_x$ (x =0.5, 1.0, 1.5, 2.0, 2.5) system.

（1）From the analysis of $AlTiNiCuCo_x$ system high entropy alloy phase, we can find that with the increase of Co element content, alloy system FCC and BCC structure changed. The Cu phase of the FCC structure in the alloy system always exists. When the molar ratio of Co element is at 1.0, 1.5, 2.0, 2.5, the new BCC structure TiCo phase and the NiTi phase disappeared and the FCC2 structure phase changed.

（2）The microstructure of $AlTiNiCuCo_x$ high-entropy alloy was analyzed by means of metallographic and SEM images. With the increase of the amount of Co element, the shape of the dendrites changed from the line filaments to the regular dendrites, and the dendrite arms began to expand laterally to the snowflake-like transition.

（3）Combined with XRD, SEM-EDS and TEM to analyze the high entropy alloy of $AlTiNiCuCo_x$ system, the intergranular region consisted of Cu-rich phase of FCC structure, the inner grain was rich in Ti, Ni, A1 and Co elements, and the inner matrix of the grain was mostly in the colattice form with the second phase.

## References

1. Senkov, O.N. and D.B. Miracle, *A new thermodynamic parameter to predict formation of solid solution or intermetallic phases in high entropy alloys*. Journal of Alloys and Compounds, 2016. **658**: p. 603-607.

2. Toda-Caraballo, I. and P.E.J. Rivera-Díaz-del-Castillo, *A criterion for the formation of high entropy alloys based on lattice distortion.* Intermetallics, 2016. **71**: p. 76-87.

3. Li, B.S., et al., *Effects of Mn, Ti and V on the microstructure and properties of AlCrFeCoNiCu high entropy alloy.* Materials Science and Engineering: A, 2008. **498**(1-2): p. 482-486.

4. Yurchenko, N., N. Stepanov, and G. Salishchev, *Laves-phase formation criterion for high-entropy alloys*. Materials Science and Technology, 2016. **33**(1): p. 17-22.

5. Tsai, M.-H., et al., *Incorrect predictions of simple solid solution high entropy alloys: Cause and possible solution.* Scripta Materialia, 2017. **127**: p. 6-9.

6. Tsai, M.-H. and J.-W. Yeh, *High-Entropy Alloys: A Critical Review.* Materials Research Letters, 2014. **2**(3): p. 107-123.

7. Cantor, B., et al., *Microstructural development in equiatomic multicomponent alloys.* Materials Science and Engineering: A, 2004. **375-377**: p. 213-218.

8. Yeh, J.-W., et al., *Formation of simple crystal structures in Cu-Co-Ni-Cr-Al-Fe-Ti-V alloys with multiprincipal metallic elements.* Metallurgical and Materials Transactions A, 2004. **35**: p. 2533-2536.

9. Chen, T.K., et al., *Nanostructured nitride films of multi-element high-entropy alloys by reactive DC sputtering.* Surface and Coatings Technology, 2004. **188-189**: p. 193-200.

10. Tong, C.-J., et al., *Mechanical performance of the AlXCoCrCuFeNi high-entropy alloy system with multiprincipal elements.* Metallurgical and Materials Transactions A-physical Metallurgy and Materials Science - METALL MATER TRANS A, 2005. **36**: p. 1263-1271.

11. Shi, Y., B. Yang, and P. Liaw, *Corrosion-Resistant High-Entropy Alloys: A Review.* Metals, 2017. **7**(2).

12. Huang, P.K., et al., *Multi-Principal-Element Alloys with Improved Oxidation and Wear Resistance for Thermal Spray Coating.* Advanced Engineering Materials, 2004. **6**(12): p. 74-78.

13. Guo, S., C. Ng, and C.T. Liu, *Anomalous solidification microstructures in Co-free AlxCrCuFeNi2 high-entropy alloys.* Journal of Alloys and Compounds, 2013. **557**: p. 77-81.

14. Huang, Y.-S., et al., *Microstructure, hardness, resistivity and thermal stability of sputtered oxide films of AlCoCrCu0.5NiFe*

*high-entropy alloy.* Materials Science and Engineering: A, 2007. **457**(1-2): p. 77-83.

15. Lin, C.H., J.G. Duh, and J.W. Yeh, *Multi-component nitride coatings derived from Ti–Al–Cr–Si–V target in RF magnetron sputter.* Surface and Coatings Technology, 2007. **201**(14): p. 6304-6308.
16. Miracle, D.B. and O.N. Senkov, *A critical review of high entropy alloys and related concepts.* Acta Materialia, 2017. **122**: p. 448-511.
17. Gente, C., M. Oehring, and R. Bormann, *Formation of thermodynamically unstable solid solutions in the Cu-Co system by mechanical alloying.* Phys Rev B Condens Matter, 1993. **48**(18): p. 13244-13252.
18. Birol, Y., *Semi-solid processing of the primary aluminium die casting alloy A365.* Journal of Alloys and Compounds, 2009. **473**(1-2): p. 133-138.
19. Koch, C.C., *Intermetallic matrix composites prepared by mechanical alloying - A review.* Materials Science and Engineering: A, 1998. **244**: p. 39-48.
20. Grant, P., *Spray forming.* Progress in Materials Science, 1995. **39**: p. 497–545.

21. Lin, C.-M. and H.-L. Tsai, *Effect of annealing treatment on microstructure and properties of high-entropy FeCoNiCrCu0.5 alloy.* Materials Chemistry and Physics, 2011. **128**(1-2): p. 50-56.

## Acknowledgement

This study is supported by the Chinese Marine Low-Speed Engine Project-Phase I with Grant No. CDGC01-KT0302.

## Author contributions statement

Junfeng Wang: Conceptualization, Methodology, Writing - review and editing, Visualization. Xinran Wang: Formal analysis, Resources. Zhihao Huang: Validation, Resources. Ting Li: Conceptualization, Formal analysis. Yudong Fu: Supervision, Methodology, Resources.

## Additional information

Declaration of competing interest

The authors declare that they have no known competing financial interests or personal relationships that could have appeared to influence the work reported in this paper.